\documentclass[onecolumn,showpacs,preprint,preprintnumbers,amsmath,amssymb]{revtex4}
\usepackage{amssymb}
\usepackage[dvips]{graphicx}
\begin{document}

\title{ Bose-Einstein condensation in the pseudogap phase of cuprate superconductors}
\author{A. S. Alexandrov}

\affiliation{Department of Physics, Loughborough University,
Loughborough LE11 3TU, United Kingdom\\}

\begin{abstract}
We have identified  the unscreened Fr\"ohlich electron-phonon
interaction (EPI)  as the most essential for pairing in cuprate
superconductors as now confirmed by isotope substitution, recent
angle-resolved photoemission (ARPES),   and  some other experiments.
Low-energy physics is that of mobile lattice polarons and bipolarons
in the strong EPI regime. Many experimental observations have been
predicted or explained in the framework of our "Coulomb-Fr\"ohlich"
model, which fully takes into account the long-range Coulomb
repulsion and the Fr\"ohlich EPI. They
 include pseudo-gaps,  unusual isotope effects and upper critical fields, the
normal state Nernst effect, diamagnetism, the Hall-Lorenz numbers,
and a giant proximity effect (GPE). These experiments along with the
parameter-free estimates  of the Fermi energy and the critical
temperature support a genuine Bose-Einstein condensation of
real-space lattice bipolarons in the pseudogap phase of cuprates. On
the contrary the phase fluctuation (or vortex)  scenario is
incompatible with the insulating-like in-plane resistivity and the
magnetic-field dependence of orbital magnetization in the resistive
state of underdoped cuprates.

\end{abstract}


\maketitle

\section{Introduction}

Relatively high superfluid $T_c$  of $^{4}He$ ($\approx 2.17 K$ )
compared with $T_c$ of $^{3}He$ ($\approx 0.0026K$) kindles the view
that high-temperature superconductivity might derive from preformed
real-space charged bosons rather than in the BCS state with the
strongly overlapping  Cooper pairs. Indeed, there is increasing
experimental evidence that cuprates are bosonic superconductors
\cite{alemot,alebook,edwards}. A possible fundamental origin of such
strong departure of the cuprates from conventional BCS behaviour is
the unscreened (Fr\"ohlich) EPI of the order of 1 eV
\cite{ale96,alebra}, routinely neglected in the Hubbard $U$ and
$t-J$ models of cuprate superconductors
 \cite{tJ}. This interaction with $c-$axis polarized optical phonons
is virtually unscreened because the upper limit for the out-of-plane
plasmon frequency ($\lesssim 200$ cm$^{-1}$ \cite{plasma}) in
cuprates is well below the characteristic frequency of optical
phonons, $\omega_0\approx$ 400 - 1000 cm $^{-1}$. Since
 screening is poor, the magnetic interaction remains
small compared with the Fr\"ohlich EPI at any doping of cuprates. In
order to generate a convincing theory of high-temperature
superconductivity, one has to treat the long-range Coulomb repulsion
and the \emph{unscreened} EPI on an equal footing. When both
interactions are strong compared with the kinetic energy of
carriers, this Coulomb-Fr\"ohlich model (CFM)  predicts the ground
state in the form of mobile small bipolarons, which bose-condense at
high temperatures \cite{ale96,alekor2,jim2}.

Most compelling evidence for (bi)polaronic carries in cuprate
superconductors is provided by  the discovery of a substantial
isotope effect on the carrier mass \cite{guo} predicted for the
(bi)polaronic conductors in Ref. \cite{aleiso}. High resolution
ARPES \cite{lan0,shen} provides another piece of evidence for a
strong EPI in  cuprates \cite{allen} apparently with
c-axis-polarised optical phonons \cite{shen}. These as well as
 earlier optical
\cite{opt}, neutron scattering \cite{neutron} and more recent
tunnelling \cite{tun} experiments unambiguously show that the
lattice vibrations  play a significant though unconventional role in
high temperature superconductors. Operating together with a shorter
range deformation potential and molecular-type (e.g. Jahn-Teller
\cite{muller}) EPIs, the Fr\"ohlich EPI readily overcomes the
Coulomb repulsion at a short distance about the lattice constant
providing a non-retarded attraction to form small  yet mobile
bipolarons \cite{alebook}.

When strong EPI  binds holes into intersite oxygen bipolarons
\cite{ref}, the chemical potential remains pinned inside the charge
transfer gap, as clearly observed in the tunnelling experiments by
Bozovic et al. in optimally doped La$_{1.85}$Sr$_{0.15}$CuO$_4$
\cite{boz0}. The bipolaron binding energy and the singlet-triplet
bipolaron exchange energy
 are thought to be the origin of normal state charge and
spin pseudogaps, respectively, as  has been predicted by us
\cite{alegap} and later supported experimentally \cite{kabmic}. In
overdoped samples carriers screen part of EPI with low frequency
phonons. Hence, the bipolaron binding energy decreases
\cite{alekabmot} and the hole bandwidth increases with doping. As a
result, the chemical potential enters the oxygen band in overdoped
samples, so that a Fermi-level crossing could be seen in ARPPES at
overdoping where mobile bipolarons  coexist with the polaronic
Fermi-liquid \cite{narlikar}.

Here I briefly review a number of theoretical and experimental
observations supporting the bosonic model of superconducting
cuprates at variance with  BCS-like and phase-fluctuation scenarios.

\section{Low Fermi energy, individual pairing and  parameter-free
evaluation of $T_c$}

A parameter-free estimate of the Fermi energy \cite{alefermi}
clearly supports the real-space (i.e individual) pairing in cuprate
superconductors.  The band structure of cuprates is
quasi-two-dimensional. Applying the parabolic approximation for the
band dispersion one readily estimates the Fermi energy in 2D  as
$\epsilon _{F}=d\hbar^2 c^2/4e^{2}\lambda _{ab}^{2}$, where $d$ is
the distance between copper-oxygen planes, $\lambda _{ab}^{-2}=4\pi
e^{2}n/m^{\ast}c^2$ is the in-plane magnetic-field penetration depth
at low temperatures, and $n,m^{\ast }$ are the density of hole
polarons and their effective mass, respectively. Here one assumes
that the `superfluid' density at zero temperature is about the same
as the normal state density of holes, as it must be in any clean
superfluid \cite{pop}. The low-temperature penetration depth is
unusually large, $\lambda_{ab}\gtrsim 150$ nm, in cuprates, so that
the renormalised Fermi energy turns out to be surprisingly low, $
\epsilon _{F} \lesssim 100$ meV, rendering the Migdal-Eliashberg
theory \cite{mig,eli} inadequate. In fact, $\epsilon _{F}$ is so
small that the individual pairing is very likely. Such pairing will
occur when the size of the pair, $r_b=\hbar/\sqrt{m^{\ast }\Delta }$
is small compared with  the inter-pair separation, $r=\hbar
\sqrt{\pi/m^{\ast }\epsilon _{F}}$, so that the condition for
real-space pairing is $\epsilon _{F}\lesssim \pi \Delta$.
Experimentally measured pseudogaps of many cuprates are about $50$
meV or larger. If one accepts that the pseudogap is about half of
the pair  binding energy, $\Delta/2$, \cite{alegap} then the
condition for real-space pairing is well satisfied in most cuprates
(typically $r_b\approx 0.2 -0.4$ nm).

When bipolarons are small so that pairs do not overlap, the pairs
can form a Bose-Einstein condensate (BEC) \cite{alebook}. Recent
Quantum Monte Carlo simulations of CFM  show that with realistic
values of EPI coupling constant, $\lambda\simeq 1$, and high optical
phonon frequencies one can avoid overlap of pairs and get a very
high bose-condensation temperature $T_{BEC}$  \cite{jim2}. Actually
$T_{BEC}$ fits the  experimental  $T_c$ in a great number of
cuprates without any  fitting parameters \cite{alekab}. In contrast
with Ref. \cite{uem} our
 expression for $T_{BEC}$   involves not only the
in-plane, $\lambda _{ab}$, but also the out-of-plane, $\lambda
_{c}$, magnetic field penetration depth, and the normal state Hall
ratio $R_{H}$ just above the transition,
\begin{equation}
T_{BEC}\approx1.64 \left( {\frac{%
eR_{H}}{{\lambda _{ab}^{4}\lambda _{c}^{2}}}}\right) ^{1/3}.
\label{Tc}
\end{equation}
Here $T_{c}$ is measured in Kelvin, $eR_{H}$ in cm$^{3}$ and
$\lambda $ in cm.  Since all quantities in Eq.(\ref{Tc}) are
measurable, the bipolaron theory provides the parameter-free
expression, which unambiguously tells us how near cuprates are to
the BEC regime. Its comparison  with the experimental $T_{c}$ of more than $%
30$ different cuprates, where both $\lambda _{ab}$ and $\lambda
_{c}$ have been measured along with $R_{H}(T_{c}+0)$, show that
 $T_{BEC}$ fits experimental values within an
experimental error bar for the penetration depths.

One can argue that due to a large anisotropy cuprates may belong to the 2D `$%
XY $' universality class with the Kosterlitz-Thouless  (KT)
superfluid critical temperature $T_{KT}$ of  the Cooper pairs
\cite{kiv}. The KT transition temperature is expressed through the
in-plane penetration depth alone as $T_{KT}\approx d\hbar
^{2}c^2/4k_B\pi e^{2}\lambda _{ab}^{2}$. It turns out significantly
higher than the experimental values in many cuprates.
There are also quite a few samples with about the same $%
{\lambda _{ab}}$ and about the same inter-plane distance $d$, but
with very different values of $T_{c},$ which makes the KT scenario
unviable. On the contrary, our parameter-free fit of the
experimental critical temperature and the critical behavior (see
below) favor $3D$ BEC of charged bosons as the mechanism of high
T$_{c}$ rather than any low-dimensional phase-fluctuation scenario.

\section{Upper critical field and the normal state Nernst effect}

  The state of bipolarons above the resistive
$T_{c}$ is perfectly "normal" in the sense that the off-diagonal
 order  parameter (i.e. the Bogoliubov-Gor'kov
anomalous average $\cal{F}(\mathbf{r,r^{\prime }})=\langle
\psi_{\downarrow }(\mathbf{{r})\psi _{\uparrow }({r^{\prime
}}\rangle}$) is zero (here $\psi_{\downarrow,\uparrow }(\mathbf{r})$
annihilates electrons with spin $\downarrow, \uparrow$ at the point
${\bf r}$).

On the contrary   $\cal{F}(\mathbf{r,r^{\prime }})$ remains nonzero
in the phase fluctuation scenario \cite{kiv} with vortexes in the
normal state  above $T_{c}$ \cite{xu,ong}. We have noticed that the
phase fluctuation scenario is not compatible with  extremely sharp
resistive transitions at $T_c$ in high-quality underdoped, optimally
and overdoped cuprates, and with the insulating-like in-plane
resistivity of underdoped cuprates in high magnetic fields
\cite{alezav,alenernst}. For example, the in-plane and out-of-plane
resistivity of $Bi-2212$, where the anomalous Nernst signal has been
measured \cite{xu}, is perfectly "normal" above $T_c$,  showing only
a few percent positive or negative magnetoresistance \cite{zavale},
explained with bipolarons \cite{mos}.

\begin{figure}[tbp]
\begin{center}
\includegraphics[angle=-0,width=0.60\textwidth]{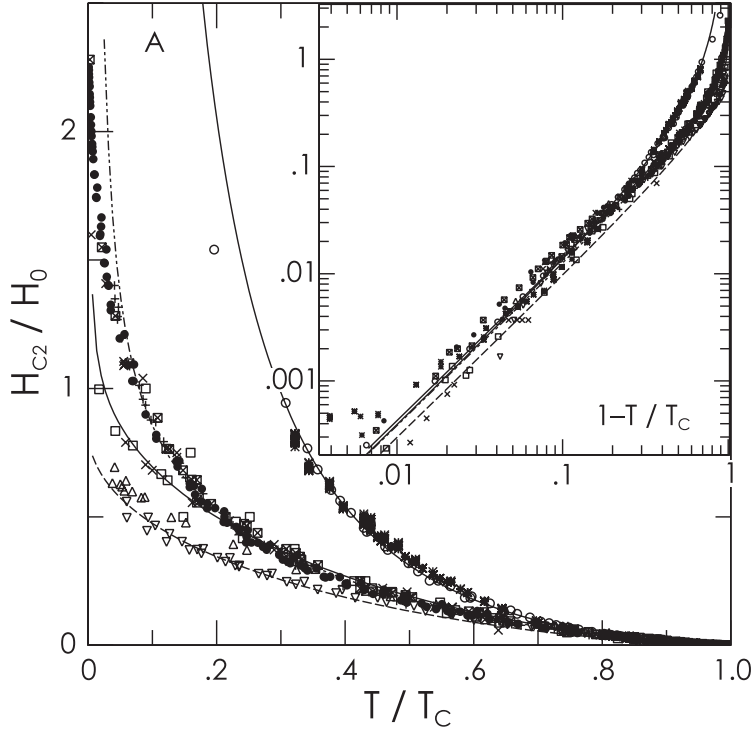} \vskip -0.5mm
\end{center}
\caption{\small{Resistive upper critical field \cite{zavkabale}
(determined at 50\% of the transition) of cuprates, spin-ladders and
organic superconductors scaled according to the Bose-Einstein
condensation field of charged bosons \cite{aleH},
$H_{c2}(T)\propto\left[b(1-t)/t+1-t^{1/2}\right] ^{3/2}$ with
$t=T/T_c$.  The parameter $b$ is proportional to the number of
delocalised bosons at zero temperature, $b$ is 1 (solid line), 0.02
(dashed-dotted line), 0.0012 (dotted line), and 0 (dashed line). The
inset shows a universal scaling of the same data near $T_{c}$ on the
logarithmic scale. Symbols correspond to $Tl-2201(\bullet) $,
$La_{1.85}Sr_{0.15}CuO_{4} (\triangle)$, $Bi-2201 (\times)$,
$Bi-2212 (\ast)$, $YBa_{2}Cu_3O_{6+x}(\circ)$,
$La_{2-x}Ce_{x}CuO_{4-y}(\square)$, $Sr_2Ca_{12}Cu_{24}O_{41} (+)$,
and Bechgaard salt organic superconductor ($\triangledown)$ }.}
\label{Hc2}
\end{figure}

Both in-plane  and out-of-plane resistive transitions  of
high-quality samples
 remain sharp in the magnetic field providing a
reliable determination of the genuine upper critical field,
$H_{c2}(T)$.
 Many high magnetic field studies revealed a non-BCS upward
curvature of  $H_{c2}(T)$ (for review see \cite{zavkabale}) with a
non-linear temperature dependence in the vicinity of $\ T_{c}$  in
many cuprates and some other unconventional superconductors,
Fig.\ref{Hc2}. If unconventional superconductors are in the
`bosonic' limit, such unusual critical fields are expected in
accordance with the theoretical prediction for BEC of charged bosons
in the external magnetic field \cite{aleH}.

Also the bipolaron theory  accounts for the anomalous Nernst signal,
and the insulating-like in-plane  resistivity of underdoped cuprates
\cite{alezav,alenernst} as observed \cite{xu,cap,cap2,ong}.
 The transverse
Nernst-Ettingshausen effect (here the Nernst effect) is the
appearance of a transverse electric field $E_y$ perpendicular to the
magnetic filed and to a temperature gradient, $\nabla _{x}T$. When
bipolarons are formed  the chemical potential is negative in the
normal state. It is found in the impurity band just below the
mobility edge at $T>T_c$. Carriers, localised below the mobility
edge contribute to the longitudinal transport together with the
itinerant carriers in extended states above the mobility edge.
Importantly the contribution of localised carriers  adds  to the
contribution of itinerant carriers to produce a large Nernst signal,
$e_{y}\equiv -E_{y}/\nabla _{x}T$, while it reduces the thermopower
$S$ and the Hall angle $\Theta$. This unusual "symmetry breaking" is
at variance with conventional metals where the familiar "Sondheimer"
cancelation \cite{sond} makes $e_{y}$ much smaller than $S\tan
\Theta$ because of the electron-hole symmetry near the Fermi level.
Such  behaviour originates in the "sign" (or "$p-n$") anomaly of the
Hall conductivity of localised carriers. The sign of their Hall
effect is often $opposite$ to that of the thermopower as observed in
many amorphous semiconductors \cite{ell}.

\begin{figure}
\begin{center}
\includegraphics[width=0.5\textwidth]{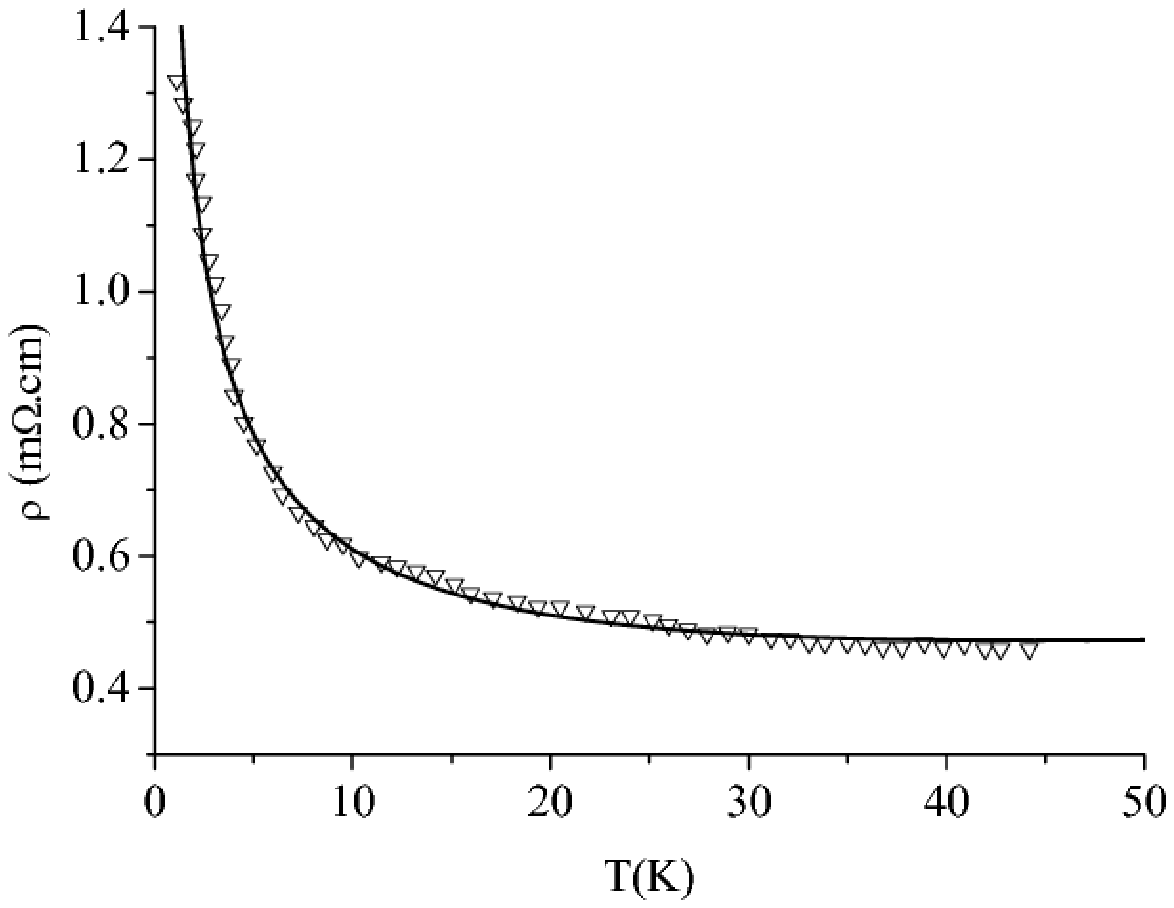}
\caption{Normal state in-plane resistivity of underdoped La$_{1.94}$
Sr$_{0.06}$CuO$_4$ (triangles \cite{cap}) as revealed in the field
$B=12$ Tesla  and compared with the bipolaron theory \cite{narlikar}
(solid line).} \label{rho}
\end{center}
\end{figure}

Hence the bipolaron model  can account for a low value of
$S\tan\Theta $ compared with a large value of $e_y$ in some
underdoped cuprates \cite{xu,cap2} due to the sign anomaly. Near the
mobility edge   $S\propto T$ as in conventional  amorphous
semiconductors \cite{mott3}, so that $S\tan \Theta  \propto
T/\rho(T)$ and $e_{y}\propto (1-T/T_1)/\rho(T)$.

According to our earlier suggestion \cite{alelog} the
insulating-like low-temperature dependence of $\rho(T)$ in
underdoped cuprates  originates from the elastic scattering of
nondegenerate itinerant carriers by charged  impurities.  The
relaxation time of nondegenerate carriers depends on temperature as
$\tau \propto T^{-1/2}$ for scattering by short-range deep potential
wells, and as $T^{1/2}$ for  shallow wells \cite{alelog}. Combining
both scattering rates one obtains $\rho
=\rho_0[(T/T_2)^{1/2}+(T_2/T)^{1/2}]$, which fits extremely well the
experimental insulating-like normal state resistivity of underdoped
La$_{1.94}$ Sr$_{0.06}$CuO$_4$, Fig. \ref{rho},  with $\rho_0=0.236$
m$\Omega\cdot$cm and $T_2=44.6$K. Then  $S\tan \Theta$ and $e_y$ can
be parameterized as
\begin{equation}
S\tan \Theta = e_0
{(T/T_2)^{3/2}\over{1+T/T_2}},
\end{equation}
and
\begin{equation}
e_{y}(T,B)=e_0{(T_1-T) (T/T_2)^{1/2}\over{T_2+T}} ,
\end{equation}
where $T_1$ and $e_0$ are temperature independent.
\begin{figure}
\begin{center}
\includegraphics[angle=270,width=0.70\textwidth]{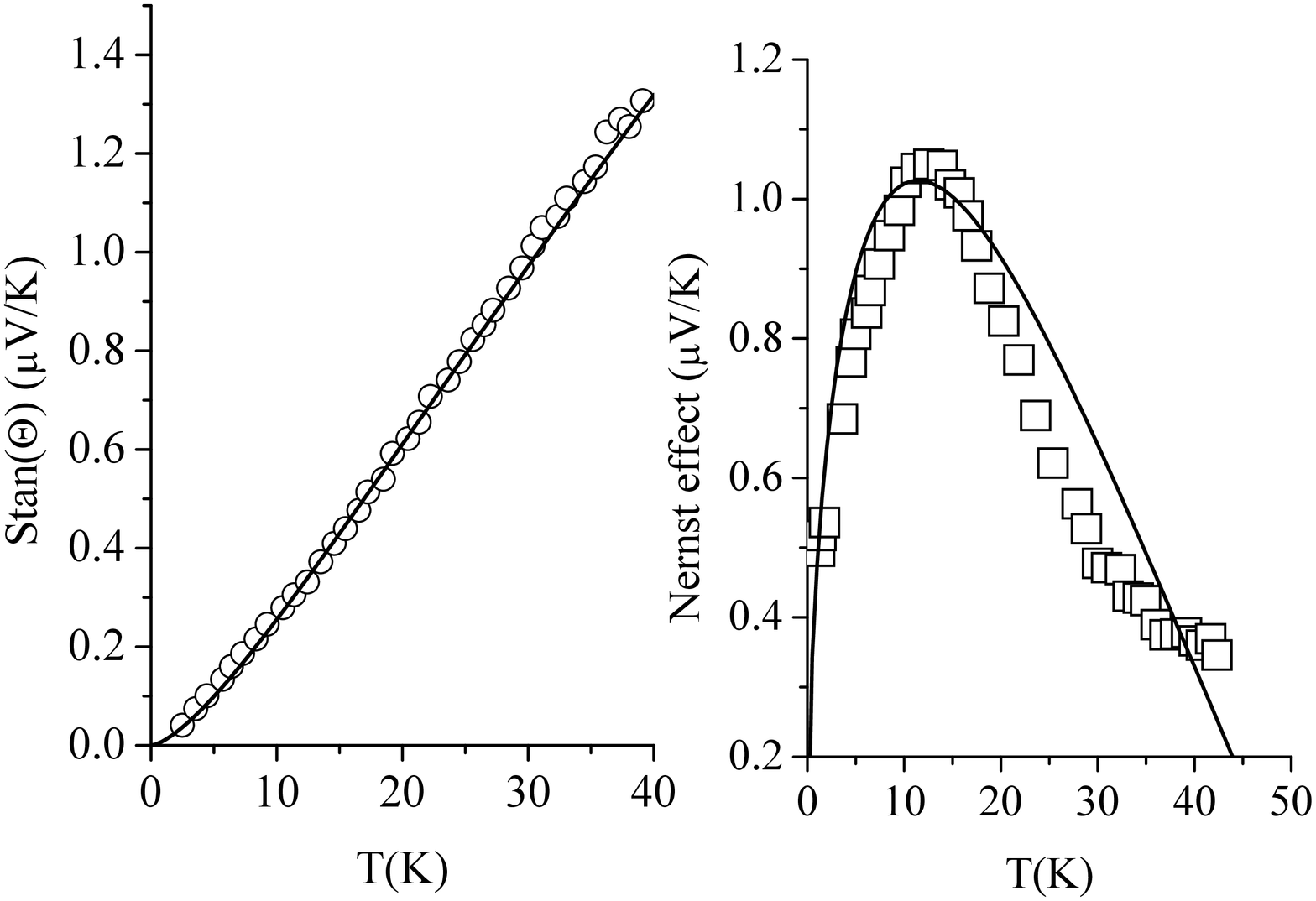}
\vskip -0.5mm \caption{$S\tan\Theta$ (circles \cite{cap2} )  and the
Nernst effect $e_y$  (squares \cite{cap})  of underdoped La$_{1.94}$
Sr$_{0.06}$CuO$_4$ at $B=12$ Tesla compared with the bipolaron
theory (solid lines) \cite{alenernst}.} \label{S}
\end{center}
\end{figure}

In spite of all simplifications, the model describes  remarkably
well both $S\tan \Theta$ and $e_y$, Fig.\ref{S},  measured in
La$_{1.94}$ Sr$_{0.06}$CuO$_4$ with a $single$ fitting parameter,
$T_1=50$K using the experimental $\rho(T)$. The constant  $e_0=2.95$
$\mu$V/K scales the magnitudes of $S\tan \Theta$ and $e_y$.  The
magnetic field $B=12$ Tesla destroys the superconducting state of
the low-doped La$_{1.94}$ Sr$_{0.06}$CuO$_4$ down to $2$K,
Fig.\ref{rho}, so any residual superconducting order above $2$K
(including vortexes) is clearly ruled out, while the Nernst signal,
 is remarkably large. The coexistence of the large Nernst
signal and a nonmetallic resistivity is in sharp disagreement with
the vortex scenario, but in agreement with our model.

\section{Hall-Lorenz number and normal state diamagnetism}

The  measurements of the Righi-Leduc effect provides further
evidence for
 charged bosons  above $T_c$ \cite{leeale}.
  The
effect describes transverse heat flow resulting from a perpendicular
temperature gradient in an external magnetic field, which is  a
thermal analog of the Hall effect. Using the effect the
"Hall-Lorenz" electronic number, $ L_{H}=\left( e/k_{B}\right)
^{2}\kappa _{xy}/(T\sigma _{xy})$ has been directly measured
 \cite{ZHANG} in $YBa_{2}Cu_{3}O_{6.95}$ and $YBa_{2}Cu_{3}O_{6.6}$
since transverse thermal $%
\kappa _{xy}$ and electrical $\sigma _{xy}$ conductivities involve
no phonons. The experimental $L_{H}(T)$ showed a quasi-linear
temperature dependence above the resistive $T_{c}$, which strongly
violates the Wiedemann-Franz (WF) law. Remarkably, the measured
value of $L_{H}$ just above $T_{c}$ turned out about the same as
predicted by the bipolaron theory \cite{NEV}, $L=0.15L_{0}$, where
$L_{0}=\pi^2/3$ is the conventional Sommerfeld value. The breakdown
of the WF law revealed in the Righi-Leduc effect \cite{ZHANG}  has
been explained by a temperature-dependent contribution of thermally
excited single polarons to the transverse magneto-transport
\cite{leeale}.

Most surprisingly the torque magnetometery  \cite{mac,nau} uncovered
a diamagnetic signal somewhat above $T_c$ which increases in
magnitude with applied magnetic field. It has been  linked with the
Nernst signal and mobile vortexes   in the  normal state of cuprates
\cite{ong}. However, apart from the inconsistences mentioned above,
the vortex scenario of the normal-state diamagnetism is internally
inconsistent.  Accepting the vortex scenario and fitting  the
magnetization data in $Bi-2212$  with the conventional  logarithmic
field dependence \cite{ong}, one obtains surprisingly high upper
critical fields $H_{c2} > 120$ Tesla and a very large
Ginzburg-Landau parameter, $\kappa=\lambda/\xi >450$  even at
temperatures close to $T_c$. The in-plane low-temperature magnetic
field penetration depth is $\lambda=200$ nm in optimally doped
$Bi-2212$ (see, for example \cite{tallon}). Hence the zero
temperature coherence length $\xi$ turns out  about  the lattice
constant, $\xi=0.45$nm, or even smaller. Such a small coherence
length rules out the "preformed Cooper pairs"  \cite{kiv}, since the
pairs are virtually not overlapped at any size of the Fermi surface
in $Bi-2212$ . Moreover the magnetic field dependence of $M(T,B)$ at
and above $T_c$ is entirely inconsistent  with what one expects from
a vortex liquid.  While $-M(B)$  decreases logarithmically at
temperatures well below $T_c$, the  experimental curves
\cite{mac,nau,ong} clearly show that   $-M(B)$  increases with the
field at and  above $T_c$ , just opposite to what one could expect
in the vortex liquid.  This significant departure from the London
liquid behavior clearly indicates that the vortex liquid does not
appear above the resistive phase transition (see also Ref.
\cite{mac}).

On the contrary the bipolaron theory, which  predicted  the unusual
upper critical field \cite{aleH}, Fig. \ref{Hc2}, quantitatively
accounts for the normal state diamagnetism of cuprates as well
\cite{aledia}. When the
 magnetic field is applied perpendicular to the copper-oxygen
plains the quasi-2D bipolaron energy spectrum is quantized as
$E_\alpha= \hbar \omega(p+1/2) +2t_c [1-\cos(K_zd)]$, where $\alpha$
comprises $p=0,1,2,...$ and  in-plane $K_x$ and out-of-plane $K_z$
center-of-mass quasi-momenta,  $t_c$ and $d$ are the hopping
integral and the lattice period perpendicular to the planes, and
$\omega=2 eB/c\sqrt{m^{\ast \ast}_x m^{\ast \ast}_y }$. The
bipolaron spectrum consists of two degenerate branches, so-called
$"x"$ and $"y"$ bipolarons  \cite{ale96} with anisotropic in-plane
bipolaron masses $m^{\ast \ast}_x$ and $m^{\ast \ast}_y$. Expanding
the Bose-Einstein distribution function in powers of
$\exp[(\mu-E)/k_BT]$ with the negative chemical potential $\mu$ one
can after summation over $p$ readily obtain
 the boson density,
\begin{equation}
n_b={2eB\over{\pi \hbar c d}} \sum_{r=1}^{\infty} I_0(2t_c r/k_BT)
{\exp[ (\mu-\hbar\omega/2 -2t_c)r/k_BT]\over{1-\exp(-\hbar\omega
r/k_BT)}},
\end{equation}
and the magnetization,
\begin{eqnarray}
&&M(T,B)=-n_b \mu_b+  {2e k_B T\over{\pi \hbar c  d}}
\sum_{r=1}^{\infty}
I_0(2t_c r/k_BT)\times \\
&&{\exp[ (\mu-\hbar \omega/2 -2t_c)r/k_BT]\over{1-\exp(- \hbar\omega
r/k_BT)}} \left({1\over{r}}-{\hbar \omega \exp(-\hbar \omega
r/k_BT)\over{k_BT[1-\exp(-\hbar \omega r/k_BT)]}}\right).\nonumber
\end{eqnarray}
Here $\mu_b= e \hbar/c\sqrt{m^{\ast \ast}_xm^{\ast \ast}_y}$,
 and $I_0(x)$ is the modified Bessel
function. At low temperatures $T \rightarrow 0$ Schafroth's result
\cite{sha} is recovered, $M(0,B)= -n_b \mu_b$. The magnetization of
charged bosons is field-independent at low temperatures. At high
temperatures, $T \gg T_c$ the chemical potential has a large
magnitude, and we can keep only the terms with $r=1$ to obtain
$M(T,B)=-n_b \mu_b \hbar \omega/(6k_BT)$ at $T \gg T_c\gg \hbar
\omega/k_B$,
 which is the familiar  Landau  orbital diamagnetism  of nondegenerate
 carriers. Here $T_c$ is the  Bose-Einstein condensation
temperature $T_{c}= 3.31 \hbar^2 (n_{b}/2)^{2/3}/(m^{\ast
\ast}_{x}m^{\ast \ast}_{y}m^{\ast \ast}_{c})^{1/3}$, with
$m_{c}=\hbar^2/2|t_{c}|d^{2}$.

\begin{figure}
\begin{center}
\includegraphics[angle=-90,width=0.70\textwidth]{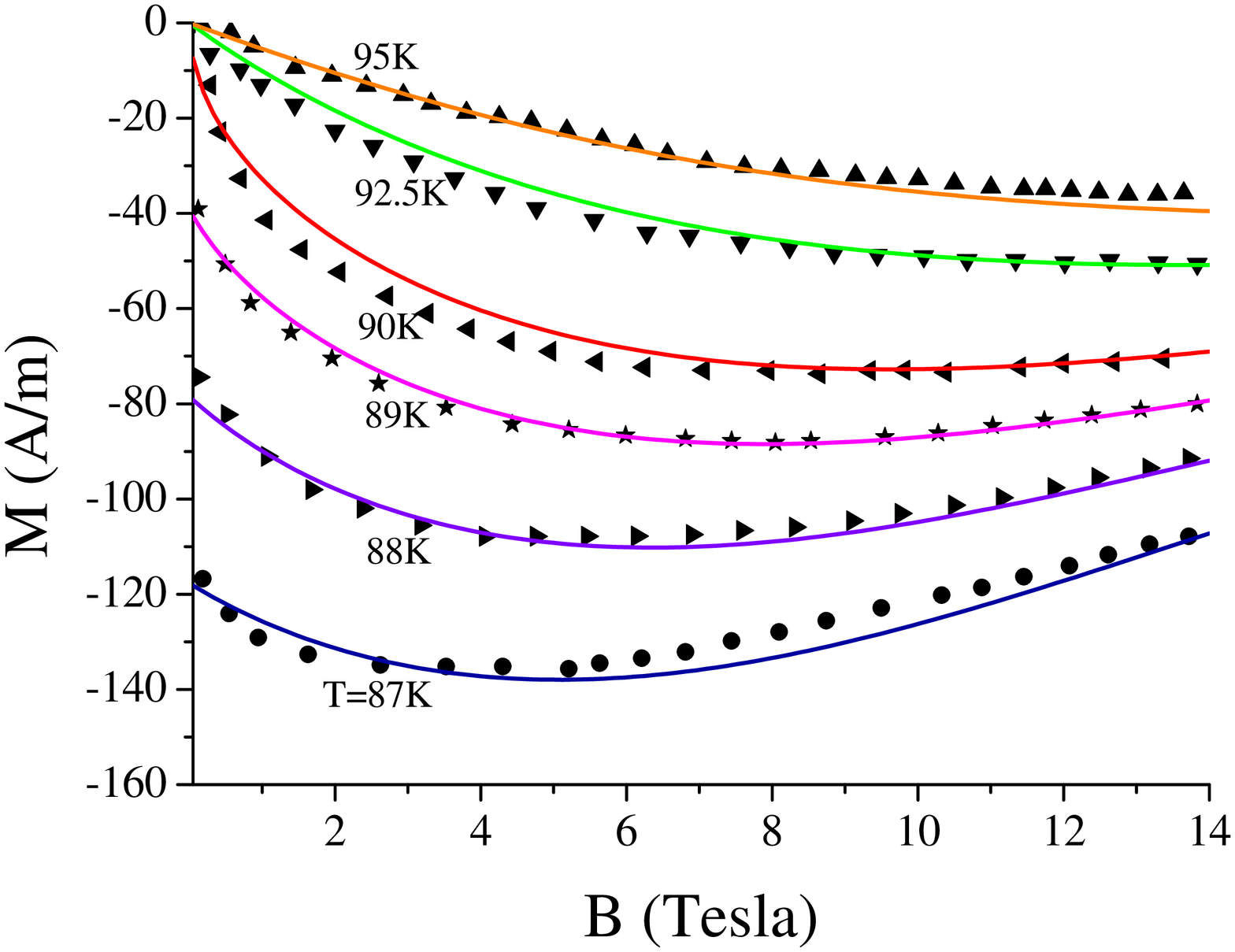}
\vskip -0.5mm \caption{Diamagnetism of optimally doped Bi-2212
(symbols)\cite{ong} compared with the magnetization of charged
bosons \cite{aledia} near and above $T_c$ (lines).} \label{dia}
\end{center}
\end{figure}

Comparing with the experimental data one has to take into account a
temperature and field depletion of singlets due to their thermal
excitations into spin-split triplets and single-polaron states by
allowing for some temperature and field dependences of the singlet
bipolaron density,
 $n_b(T,B)=n_c[1-\alpha \tau -(B/B^*)^2]$ (here $\tau=T/T_c-1$).
 As a result,  $M(T,B)$ of charged bosons fits remarkably well the  experimental
 orbital magnetization
  of  optimally doped Bi-2212, Fig. \ref{dia},   with $n_c \mu_b=
2100$A/m, $T_c=90$K,  $\alpha=0.62$ and $B^*=56$ Tesla.

On the other hand the experimental data, Fig. \ref{dia}, are in
disagreement with the phase-fluctuation scenario \cite{kiv,ong}.
Indeed, a critical exponent  $\delta=\ln B/\ln|M(T,B)|$ for
$B\rightarrow 0$, in the charged Bose gas  is dramatically different
from the KT  critical exponent. While the KT $\delta$  is about $1$
above $T_{KT}$, charged bosons have strongly temperature dependent
$\delta(T) >1$, which  is very close to the experimental $\delta(T)$
\cite{aledia}.

\section{Giant proximity effect} Several groups reported that in
the Josephson cuprate \emph{SNS} junctions  supercurrent can run
through normal \emph{N}-barriers as thick  as 10 nm or even thicker
in a strong conflict with the standard theoretical picture. Using
the advanced molecular beam epitaxy, Bozovic \emph{et al.}
\cite{bozp} proved that this giant proximity effect (GPE) is
intrinsic, rather than  caused by any inhomogeneity of the barrier
such as stripes, superconducting "islands", etc.. Hence GPE defies
the conventional explanation, which predicts that the critical
current should exponentially decay with the characteristic length of
about the coherence length, which is $\xi \lesssim 1$ nm in the
cuprates.

This unusual effect can be broadly understood as the Bose-Einstein
condensate  tunnelling into
 a cuprate \emph{semiconductor} \cite{pro}. The condensate wave function, $\psi(z)$,  is described by
the Gross-Pitaevskii (GP) equation.   In the superconducting region,
$z<0$, near the $SN$ boundary at $z=0$, the GP equation is
\begin{equation}
{\hbar^2\over{2m^{**}}}{d^2\psi(z)\over{dz^2}}=[V
|\psi(z)|^2-\mu]\psi(z), \label{psi}
\end{equation}
where $V$ is a short-range repulsion of bosons, and $m^{**}$ is the
boson mass along the direction of tunnelling $z$.  Deep inside the
superconductor $|\psi(z)|^2=n_s$ so that $\mu=Vn_s$ , where the
condensate density $n_s$ is about half of the hole density, if the
temperature is well below $T_c$. Then the coherence length  is $\xi=
\hbar/(2m^{**} n_s V)^{1/2}$ from Eq.(\ref{psi}).

If the normal barrier, at $z>0$, is an underdoped cuprate
semiconductor above its transition temperature,  the chemical
potential $\mu$ lies below the quasi-2D bosonic band by some energy
$\epsilon$ given by
\begin{equation}
\epsilon(T)\leqslant -k_BT\ln(1-e^{-T_0/T}), \label{eps}
\end{equation}
which is exponentially small at $T'_c < T \ll T_0$ turning into zero
at $T=T'_c$. Here  $T'_c \approx T_0/\ln(k_BT_0/2t_c)$, is the
transition temperature of the barrier, and $T_0=\pi \hbar^2 n'd/k_Bm
\gg T'_c \gg t_c/k_B$. The GP equation in the barrier is written as
\begin{equation}
{\hbar^2 \over{2m^{**}}}{d^2\psi(z)\over{dz^2}} =[V
|\psi(z)|^2+\epsilon]\psi(z).
\end{equation}
It predicts  the occurrence of a new length scale,
$\hbar/\sqrt{2m^{**}\epsilon (T)}$.  In a wide temperature range
 $T'_c <T < T_0$,  this length
turns out much larger than the zero-temperature coherence length,
$\xi$, since $\epsilon(T)$ in Eq.(\ref{eps}) could be very small,
which explains GPA.
 The physical reason why
the quasi-2D bosons display a large normal-state coherence length,
whereas 3D Bose-systems (or any-D Fermi-systems) at the same values
of parameters do not, originates in the large density of states
(DOS) near the band edge of two-dimensional bosons compared with 3D
DOS. Since DOS is large, the chemical potential is pinned near the
edge with the exponentially small  magnitude, $\epsilon (T)$, at
$T<T_0$. Importantly if the
 barrier is undoped ( $n' \rightarrow
 0$)  $\epsilon(T)$ becomes
 large, $\epsilon(T) \varpropto \ln(1/n')\rightarrow \infty$, for any finite temperature
 $T$.
In this case the current should exponentially decay with the
characteristic length  smaller that $\xi$, as is also experimentally
observed  \cite{boz0}.

\section{Summary}

 A first proposal for
high temperature superconductivity, made by Ogg Jr in 1946
\cite{ogg}, already involved  real-space pairing of individual
electrons into bosonic molecules  with zero total spin. This idea
was further developed as a natural explanation of conventional
superconductivity by Schafroth \cite{sha}  and Butler and Blatt
\cite{blatt}. The Ogg-Schafroth picture was practically forgotten
because it neither accounted quantitatively for the critical
behavior of conventional superconductors, nor did it explain the
microscopic nature of attractive forces which could overcome the
Coulomb repulsion between two electrons  constituting a
 pair. On the contrary highly successful for low-$T_c$ metals and alloys
the BCS theory, where two electrons  were indeed  correlated, but at
a very large distance of about $10^{3}$ times of the average
inter-electron spacing, led many researchers to believe that cuprate
superconductors should also be  "BCS-like" (maybe with strong phase
fluctuations).

However,  by extending the BCS theory towards the strong EPI, a
charged Bose liquid of small bipolarons was predicted by us
\cite{aleran} with a further prediction  that high $T_c$ should
exist in the crossover region of the EPI strength from the BCS-like
to bipolaronic superconductivity \cite{ale0}. Later on we have shown
that the unscreened Fr\"ohlich EPI combined with the strong Coulomb
repulsion provides  superlight small bipolarons, which are several
orders lighter than the Holstein bipolarons
\cite{ale96,alekor2,jim2}. The bipolaron theory predicted such key
features of cuprate superconductors as anomalous upper critical
fields \cite{aleH}, spin and charge pseudogaps \cite{alegap}, and
anomalous isotope effects \cite{aleiso}
 later observed experimentally (for review see
 \cite{alemot,alebook}). The  strong EPI has been
 experimentally established in cuprates beyond any reasonable doubt.

 We
believe that the following conditions are responsible for the
high-temperature superconductivity in cuprates \cite{jim2}:  (a) The
parent compounds are  ionic insulators with light oxygen ions to
form high-frequency optical phonons, (b) The structure is quasi
two-dimensional to ensure poor screening of EPI with c-axis
polarized phonons, (c) There is strong on-site
 Coulomb repulsion to form  intersite mobile bipolarons rather than strongly localised on-site pairs,
 (d) There are moderate carrier densities to keep the
system of small bipolarons close to the dilute regime.

Here I have shown that the  normal-state diamagnetism, the Nernst,
thermal Hall and giant proximity effects provide further evidence
for a genuine Bose-Einstein condensation of real-space lattice
bipolarons in cuprates.

 I would like  to thank  Peter Edwards,  Jim Hague, Viktor
Kabanov, Pavel Kornilovitch, John Samson and Vladimir Zavaritsky for
collaboration and  helpful discussions, and to acknowledge EPSRC
(UK) (grant numbers EP/C518365/1 and EP/D07777X/1).

\label{lastpage-01}

\end{document}